\documentclass[a4paper,twoside]{article}
\newcommand{\affil}[1]{$^{\rm #1}$}
\usepackage[authoryear]{natbib}
\bibpunct{(}{)}{;}{a}{}{,}
\usepackage{graphicx}
\date{} %Please leave the date blank
\title{\large\bf\flushleft Damn You, Little h! (or, Real-World Applications Of The Hubble Constant Using Observed And Simulated Data)}
\author{\parbox{\textwidth}{\flushleft
\vspace{-0.5cm}
{\it Darren J. Croton\affil{A,B}}\\
\vspace{0.4cm}
{\small \affil{A}\,Centre for Astrophysics \& Supercomputing, Swinburne University of Technology, PO Box 218, Hawthorn, VIC 3122, Australia}\\
{\small \affil{B}\,Email: dcroton@astro.swin.edu.au}}}
\begin{document}
\twocolumn[
\begin{changemargin}{.8cm}{.5cm}
\begin{minipage}{.9\textwidth}
\vspace{-1cm}
\maketitle
%
%
%%%%%%%%%%%%%     ABSTRACT    %%%%%%%%%%%%%
%Abstract of no more than 200 words here.
\small{\bf Abstract:} The Hubble constant, $H_0$, or its dimensionless equivalent, ``little $h$'', is a fundamental cosmological property that is now known to an accuracy better than a few percent. Despite its cosmological nature, little $h$ commonly appears in the measured properties of individual galaxies. This can pose unique challenges for users of such data, particularly with survey data. In this paper we show how little $h$ arises in the measurement of galaxies, how to compare like-properties from different datasets that have assumed different little $h$ cosmologies, and how to fairly compare theoretical data with observed data, where little $h$ can manifest in vastly different ways. This last point is particularly important when observations are used to calibrate galaxy formation models, as calibrating with the wrong (or no) little $h$ can lead to disastrous results when the model is later converted to the correct $h$ cosmology. We argue that in this modern age little $h$ is an anachronism, being one of least uncertain parameters in astrophysics, and we propose that observers and theorists instead treat this uncertainty like any other. We conclude with a ``cheat sheet'' of nine points that should be followed when dealing with little $h$ in data analysis.
 
%%%%%%%%%%%%%     KEYWORDS    %%%%%%%%%%%%%
\medskip{\bf Keywords: cosmology --- galaxies --- methods: theory --- methods: observational}
% Please write all keywords in lower case. PASA uses the
% standard list of subject headings adopted by The Astrophysical Journal
% and available from http://www.journals.uchicago.edu/ApJ/keywords_text.html.
% Keywords are separated by em-dashes, i.e. ---

%%%%%%%%DO NOT EDIT%%%%%%%%%%%%
\medskip
\medskip
\end{minipage}
\end{changemargin}
]
\small
%%%%%%%%EDIT FROM HERE%%%%%%%%%%%%

\section{Introduction}

By and large, cosmology remains a science built on phenomenology. Although an increasingly accurate model of the Universe has been determined from increasingly accurate observations \citep[e.g.][]{Blake2011, Hinshaw2012, Sanchez2013, Planck2013}, the physics of the underlying cosmological model is still yet to be understood. Hence, the favoured model of the Universe is parameterised. One parameter, the Hubble constant $H_0$, has been the focus of much attention given its importance in quantifying the expanding nature of space--time. $H_0$ (commonly refereed to by its alter ego, ``little $h$'', defined below) often appears in the measurement of galactic properties at cosmological distances, and more generally throughout computational cosmology. Its presence is required whenever an assumption about the underlying cosmology must be made, no matter how subtle. These assumptions are often hidden from the final results that feature in published research. 

This dependence, and its transparent nature, has the potential to create problems. In particular, one must know the value of little $h$ to make quoted results meaningful. Specifically, because little $h$ is a measurement-dependent quantity, different ways to measure the same property may result in different little $h$ dependencies. Comparing results can be an exercise in frustration if the terminology is different between sub-disciplines, or has changed with time.

Hence, it is perhaps not surprising that many (silently) struggle when using little $h$ in their particular scientific situation. Issues often arise when presenting results for galaxies at large distances for a particular value of $H_0$, or comparing two observations that have assumed different $H_0$ values, or making comparisons between theory and observation, where $H_0$ sometimes manifests itself differently, to name but a few examples. 

The aim of this paper is to clarify what little $h$ is and how it arises in the determination of both the observed and theoretical properties of galaxies. In particular, we examine how to compare observations and simulations that have assumed different (or no) Hubble constant values. We conclude by providing a cheat sheet that gives clear direction on the use of $H_0$ for common applications.

\section{The Origin Of Little h}

\begin{figure*}
\begin{center}
\includegraphics[scale=0.6, bb = 10 50 535 355, clip]{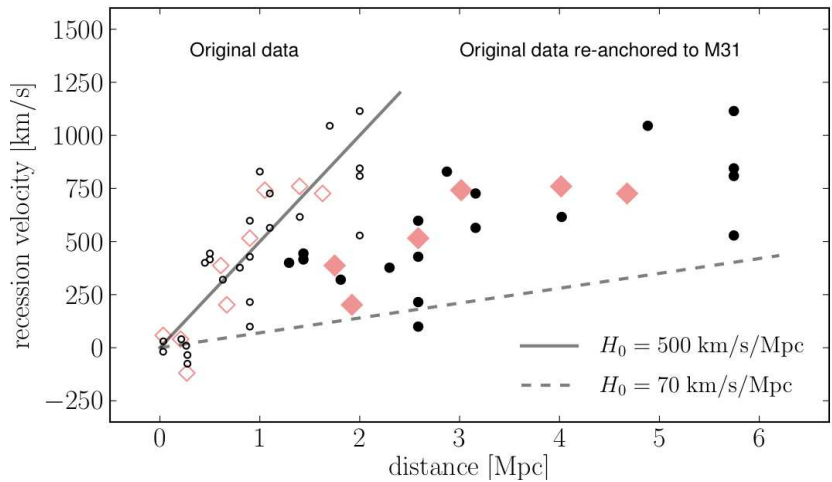}
\caption{Hubble's original measurement of distant galaxies \citep{Hubble1929}, plotting their redshift against distance. The left side shows Hubble's published data (open symbols): black circles mark individual galaxies, red diamonds group these galaxies into associations, while the solid line is his fit to the relation $v = H_0\, d$, where $H_0 = 500\, \mathrm{km}$ $\mathrm{s}^{-1}$ $\mathrm{Mpc}^{-1}$. However, the distances Hubble used are now know to be wrong. Following \citet{Peacock2013}, we re-anchor Hubble's distance ladder to the correct value of M31 and replot the data on the right side (closed symbols). The dashed line shows Hubble's Law assuming $H_0 = 70\, \mathrm{km}$ $\mathrm{s}^{-1}$ $\mathrm{Mpc}^{-1}$, close to the modern value. Clearly there were greater problems with Hubble's original distances than just its local calibration.
}
\label{fig:hubble}
\end{center}
\end{figure*}

In astronomy, everything in the Universe is moving relative to everything else. The Earth moves around the Sun, the Sun around the Milky Way, the Milky Way moves relative to the other Local Group galaxies, and the Local Group relative to more distant galaxies and galaxy clusters. Within the large-scale cosmic web we find bulk motions in every direction on the sky. Such relative velocity can be measured using a variety of techniques depending on the objects of interest. For galaxies, this is typically achieved through the identification of known spectral features which shift from where they should be because of their relative motion. The degree of this shift is known as either redshift (for galaxies moving away from us) or blueshift (for galaxies moving towards).

It was then a curious set of observations in the early 1900's that revealed that the majority of objects outside our own galaxy (then called nebulae, now known as other galaxies) were all moving away from us (i.e. redshifted), and in approximate proportion to their distance, called the distance--redshift relation. This was explicitly seen in the pioneering work of \citet{Slipher1917}, \citet{Lundmark1924}, and \citet{Hubble1929}, amongst others of the time. Although Hubble is often credited with its discovery, closer examination shows a more complex history with no one single eureka moment \citep[see][]{Peacock2013}. However, suffice to say that once the correlation was established, its ramifications changed our understanding of the Universe. 

This relationship has since come to be known as Hubble's Law, written as:
\begin{equation} \label{eqn:hubble}
v = H_0\, d~,
\end{equation}
where $v$ is the recession velocity of the galaxy, $d$ is its proper distance, and the proportionality constant, $H_0$ (Hubble's constant), was determined by fitting the data and has units of inverse time. Hubble estimated $H_0\, \sim 500\, \mathrm{km}$ $\mathrm{s}^{-1}$ $\mathrm{Mpc}^{-1}$, as shown on the left-hand side of Figure~\ref{fig:hubble} (open symbols and solid line), where we have reproduced the data and fit from his original paper (we will discuss the closed symbols on the right-hand side below). 

The observed ``fact'' that every distant galaxy in the Universe appears to be red and not blueshifted is itself remarkable. In effect, it tells us that the motions of all galaxies beyond our local volume are in a direction away from us, and the Hubble diagram shows that the further away a galaxy is, the faster its recession velocity. This was, of course, predicted by a simple solution to Einstein's equations of general relativity assuming a Friedman-Lema\^{i}tre-Robinson-Walker metric, where the scale factor was shown to have a time dependence. A somewhat crazy idea when first discovered, Einstein himself was unsatisfied with the concept of a dynamic space-time, which led him to add his famous cosmological constant, $\Lambda$. However, that is a story for another time (in a slightly expanded Universe).

During the second half of the 1900's debate raged (primarily between de Vaucouleurs and Sandage) as to the precise value of $H_0$, with observations placing it either close to 50 or 100 $\mathrm{km}$ $\mathrm{s}^{-1}$ $\mathrm{Mpc}^{-1}$ \cite[see the excellent review by][ and references therein]{Tammann2005}. The significant difference here with Hubble's original measurement was due to errors in the earlier distance calibrations (see below). The modern value of $H_0$, measured to about $2\%$ accuracy, is $67.3 \pm 1.2\, \mathrm{km}$ $\mathrm{s}^{-1}$ $\mathrm{Mpc}^{-1}$ \citep{Planck2013}. This means that a galaxy 1 Mpc away from us is receding with a velocity of 67.3 $\mathrm{km}$ $\mathrm{s}^{-1}$ due to expanding space, a galaxy at 2 Mpc is receding at 134.6 $\mathrm{km}$ $\mathrm{s}^{-1}$, and so on.

Returning to Figure~\ref{fig:hubble}, following \citet{Peacock2013} we create a modern version of this iconic plot by re-anchoring Hubble's distance ladder using the known distance to M31 of $0.79$ Mpc, then re-scale the data appropriately. This is shown on the right-hand side with closed symbols. Also shown is a fiducial Hubble Law assuming $70\, \mathrm{km}$ $\mathrm{s}^{-1}$ $\mathrm{Mpc}^{-1}$ (dashed line). The lingering disagreement reveals Hubble's distance problems ran deeper than simply calibration uncertainties. As \citet{Peacock2013} argue, he was perhaps was somewhat fortunate to be able to demonstrate any distance--redshift relation given the data he had on hand at the time.

Regardless, measurement (and later confirmation) of the Hubble expansion heralded in the age of modern cosmology. It underpins our modern cosmological paradigm. It factors in to all observed galaxy properties that need to assume a cosmology to be measured. In short, it is important and must be understood.

\section{Little h Defined}

Hubble's Law, as given by Equation~\ref{eqn:hubble}, describes the relationship between the recession velocity and distance of a galaxy. For practical application Hubble's constant is often re-expressed as
\begin{equation} \label{eqn:H0}
H_0 = 100\, h\, \mathrm{km}\,\mathrm{s}^{-1} \mathrm{Mpc}^{-1} ~,
\end{equation}
where $h$ is the dimensionless Hubble parameter, pronounced ``little $h$''. 

The subscript ``0'' in $H_0$ indicates a measurement at the present epoch and sets the normalisation of Hubble's Law. However in general, the value of the Hubble ``constant'' actually depends on redshift. Measured at redshift $z$ by an observer at redshift zero, 
\begin{equation} \label{eqn:H0_z}
H(z) = H_0\, E(z) ~,
\end{equation}
where 
\begin{equation}
E(z) = \sqrt{\Omega_\mathrm{M}(1+z)^3 + \Omega_\mathrm{k}(1+z)^2 + \Omega_\Lambda} ~.
\end{equation}
$H$(z), often called the Hubble parameter, will change with time in different ways depending on the mass ($\Omega_\mathrm{M}$), curvature ($\Omega_\mathrm{k}$) and dark energy ($\Omega_\Lambda$) densities of the Universe. However $H_0$ does not change, and thus by construction, neither does little $h$. In other words, \emph{little $h$ is just a number, a constant}\footnote{Neither is little $h$ a unit, although it is often (unfortunately) written and used like one. We will return to this point in Section~\ref{sec:summary}.}. Remember this and repeat it to everyone you meet.

Note that the units of $H_0$ are inverse time. Hence, the inverse of the Hubble constant has come to be known as the Hubble time, $t_H$:
\begin{equation}
t_H\, \equiv\, 1/H_0\, =\, 9.78\, h^{-1}\, \mathrm{Gyr} ~.
\end{equation}
And the speed of light, $c$, times the Hubble time is just the Hubble distance, $D_H$:
\begin{equation}
D_H\, \equiv\, c \times t_H\, =\, 3.00\, h^{-1}\,\mathrm{Gpc} ~.
\end{equation}
These are both fundamental numbers that astronomers often use as yardsticks against which to judge the relative age or distance of various cosmological properties. For further discussion on this and more we recommend the excellent paper, \cite{Hogg1999}.

\section{How Little h Arises In The Measurement Of Galaxies}
\label{sec:measure}

The Hubble constant is a global cosmological property of the Universe (at least in the vanilla $\Lambda$CDM paradigm) and not a local property of the individual objects within it (galaxies, gas, ...). However, \emph{measurement} of such objects does often depend on the background cosmology, and thus uncertainties about that background can propagate into the measurements themselves. In this way, the Hubble constant (typically expressed as the dimensionless little $h$ parameter) can appear in the quoted values of galaxy (and other) properties. Significantly, from Equation~\ref{eqn:H0_z}, the fact that $H_0$ is separable from $E(z)$ is the reason why we can separate out little $h$ in such measured quantities, even when the dynamics of the background cosmology are complicated and evolve.
 
Let us take a common but simple example from extragalactic astronomy. Observationally, the measurement of a galaxy's stellar mass often carries a $h^{-2}$ dependence. So, picking a random example galaxy, its mass might be written as $M_\mathrm{stars}\, =\, 10^{10.5} h^{-2} M_\odot$. 
This particular $h$ dependence arises from the way galaxy masses can be determined from the light that the telescope collects. In short, a galaxy's luminosity is drawn from the observed apparent magnitude and thus flux, the latter of which has units of area. Area has units of distance squared, for which each dimension carries the little $h$ uncertainty through the angular diameter distance. Thus, galaxy luminosity carries an inverse $h$ squared dependence. Moving from luminosity to stellar mass is non-trivial (one way is to multiply the luminosity by a stellar population model mass-to-light ratio). But throughout such calculations the $h^{-2}$ dependence usually carries through. 

In general, the key point to take away is that a galaxy property may or may not have a little $h$ dependence, depending on how the property was measured. The measurement itself will determine how little $h$ manifests in the property. Importantly, the numerical component of a property alone is not the value of the property. The value of the property is the combination of the number and the little $h$ uncertainty (if one exists).

\begin{figure*}
\begin{center}
\includegraphics[scale=0.6, bb = 10 50 535 350, clip]{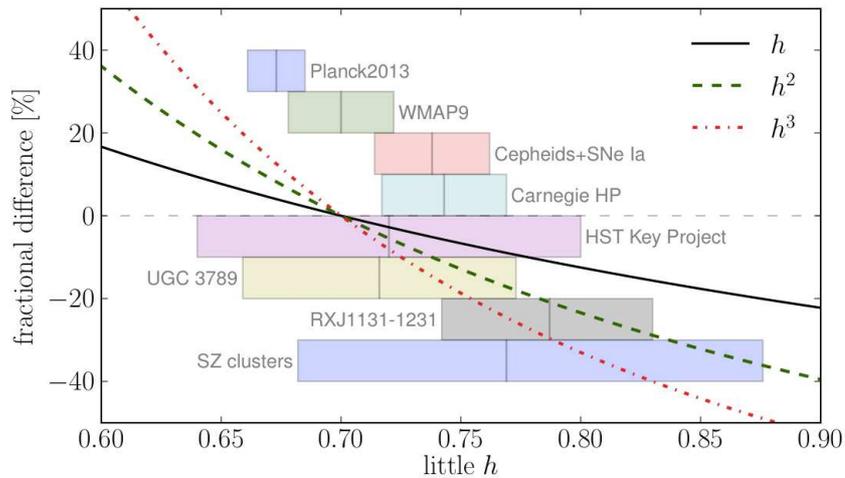}
\caption{To get a sense of the uncertainty in a property or result between two little $h$ cosmologies, we show the fractional change in the property, relative to $h=0.7$, when little $h$ is changed continuously from $0.60$ to $0.90$ (this brackets the currently favoured range). The three lines show the change for little $h$ dependencies of $h$, $h^2$ and $h^3$, as marked in the legend. Different measurements of the Hubble constant from the literature are highlighted by the shaded regions ($\pm 1 \sigma$, and spaced arbitrarily along the y-axis for clarity), taken from Figure~16 of the Planck 2013 XVI results paper \citep[][ and references therein]{Planck2013}. 
% They are (top-to-bottom): \citet{Planck2013}, ... .
}
\label{fig:fracdiff}
\end{center}
\end{figure*}

\section{How Little h Is Presented In The Literature}
\label{sec:cases}

The above all sounds simple, right? If everyone presented their results in the same way it would be. But that's not what happens in the real world. Here we identify four broad cases outlining how astronomers have dealt with little $h$ in the literature. Examples are given in Section~\ref{sec:examples}.

\begin{itemize}

\item[] \textbf{CASE~1}: The authors do not mention the chosen $H_0$ value in the paper, nor do they mention any of the $h$ dependencies when the properties of galaxies are plotted or results given. Somehow we are magically meant to know what they assumed. Perhaps there was a standard practice back when the paper was written, but alas, that lore is long lost.

\item[] \textbf{CASE~2}: The authors thankfully provide the chosen $H_0$ value near the start of the paper (usually at the end of the introduction, or in the method section), and continue to assume this value throughout the paper. However, they omit all references to $h$ when presenting their results and figures. To convert results between assumed Hubble parameter values you would either need to guess the $h$ dependency (so hopefully it's a common and obvious one) or contact the authors for clarification.

\item[] \textbf{CASE~3}: The authors mention the chosen $H_0$ value near the start of the paper and then continue to assume this value throughout the paper, as per Case~2. However, unlike Case~2, when plotting figures and discussing results they continue to explicitly show all $h$ dependencies with a subscript stating the chosen $H_0$ value (e.g. $h^{-2}_{70}$). This notation can potentially mean a number of things though, which we will illustrate below, so be warned.

\item[] \textbf{CASE~4}: The authors do not choose a $H_0$ value when presenting results and figures. Rather, properties that depend on little $h$ have had it factored out and are labeled so that the dependence is clear. Such results are numerically equivalent to a cosmology where $h=1$. Converting to your preferred $H_0$ cosmology is as simple as replacing little $h$ with the desired value and evaluating.

\end{itemize}

\subsection{Examples}
\label{sec:examples}

At this point we do not claim to make a judgment on the ``right'' way to present little $h$ in a published work (see Section~\ref{sec:summary} for that). However, it is worth emphasising how inconsistent the literature can be and why it can be a bit of a jungle for the $h$-inexperienced. We undertook an (admittedly highly incomplete) review of some of the recent highly cited literature (many of these papers have several hundred citations) to find the following examples of the above four cases:

$\bullet$ \citet{Hu2004} manage to plot the luminosity function of $z=6$ galaxies without revealing the cosmology they had assumed when converting apparent to absolute magnitude. Similarly, \citet{Shapiro2010} examine star formation in early-type galaxies using SAURON data and also compare with galaxy formation models, but fail to mention the cosmology they had adopted. These are two recent examples of Case~1 above.

$\bullet$ Two good examples of Case~2, i.e. stating the cosmology early then dropping little $h$ for the rest of the paper, are \citet{Schawinski2010} and \citet{Peng2010}. The first examines AGN and their host galaxies using Galaxy Zoo and SDSS data. The second studies the mass function with SDSS and zCOSMOS data. Both are quite clear in their application. There are countless other examples of this common usage.

$\bullet$ Case~3 and its variants can be a lot of fun. The most popular is to define $h_{70} \equiv H_0/70 = 1.0$, assuming $H_0 = 70\, \mathrm{km}\,\mathrm{s}^{-1}$ $\mathrm{Mpc}^{-1}$, as e.g. \citet{Cooray2006} did. Then, all presentations of $h_{70}$ are mathematically neutral, with the chosen $H_0$ already absorbed into the numerical value of the properties being presented. However variations can and do crop up in the literature, often unintentionally. For example, \citet{Maughan2006} mistakenly\footnote{Private communication.} define $h_{70} \equiv H_0/100 = 0.7$, while \citet{Hildebrandt2009} break all the rules and claim $h = 100/H_0$ in their paper. \citet{Drory2005} also take $H_0 = 70\, \mathrm{km}\,\mathrm{s}^{-1}$ $\mathrm{Mpc}^{-1}$ and quote $h_{70}$ when presenting results, but do not clarify its exact definition.

% For example, \citet{Maughan2006} define $h_{70} \equiv H_0/100 = 0.7$, while \citet{Cooray2006} define $h_{70} \equiv H_0/70 = 1.0$, even though both authors assume $H_0 = 70\, \mathrm{km}\,\mathrm{s}^{-1}$ $\mathrm{Mpc}^{-1}$ throughout their work. Here, in the first case the little $h$ notation represents a value of $0.7$, while in the second (which appears to be the most common usage) it represents a value of $1.0$. To rub salt into the wound, \citet{Drory2005} also take $H_0 = 70\, \mathrm{km}\,\mathrm{s}^{-1}$ $\mathrm{Mpc}^{-1}$ and quote $h_{70}$ when presenting their results and figures, but fail to say which of the above two definitions they mean. 

$\bullet$ Probably the most common usage of little $h$ is to factor it out and explicitly state the dependence, as per Case~4. For example, \citet{Brown2007} plot the evolution of the red galaxy luminosity function in a $h$'less universe, clearly stating how $h$ arises for each property considered. \citet{Croton2005} do the same but for galaxies in differing environments. 

Combinations of Cases~2--4 \emph{in the same paper} can also be found. Here are a few highlights:

$\bullet$ When analysing survey data, a popular trend is often to quote distances using Case~4 (e.g. $h^{-1}$ $\mathrm{Mpc}$) but absolute magnitudes using Case~2, taking $h=1$ and dropping the $h$ scaling (i.e. dropping $-5\log(h)$). See, for example, \citet{Zehavi2011}, \citet{Coil2008} and \citet{Hogg2004}\footnote{In fact, \citet{Hogg2004} don't actually state they've assumed $h=1$ for their absolute magnitudes.}. The rationale is presumably that these are equivalent representations of the data, but they are not\footnote{This is only true in a universe where $h$ actually equals 1, which is not our Universe.}. See Section~\ref{sec:theory} for some of the problems this can lead to.

% $\bullet$ Along the same lines, it can be confusing when the authors choose a Hubble constant at the start of their paper, e.g. $H_0 = 70\, \mathrm{km}\,\mathrm{s}^{-1}$ $\mathrm{Mpc}^{-1}$ (Case~2 or 3), but also factor out the $h$'s when the results are shown (Case~4). See, for example, ... . Since $h$ is a number and not a unit, are you still meant to evaluate each property assuming the stated $h$ if you want the results in this cosmology? Or have the numerical values already been evaluated assuming this $h$? In the latter case, the shown $h$ factors would be incorrect baggage\footnote{This would be like a nuclear physicist evaluating the electron charge in a calculation but still showing ``$e$'' in the result.}. In the former, what is the purpose of stating a Hubble constant at the start if it's not used?

$\bullet$ Another example is the famous ``NFW'' paper by \citet{Navarro1997}. The assumed cosmologies for their simulations are clearly stated in their Section~2.1, and halo masses are discussed using $h^{-1}\, M_\odot$ in the subsequent text. However, Figures~2 (showing density) and 7 (showing mass) then make no reference to little $h$, whereas they should if a consistent terminology is being employed. Are we to assume Case~2 or 4 here if we want to compare with their results?

$\bullet$ It is common to compare model galaxy stellar mass functions with observations. However, while observed mass often caries a $h^{-2}$ dependence, simulated mass only caries a $h^{-1}$ dependence, as we will discuss in Section~\ref{sec:theory}. It is curious then to see stellar mass function comparisons showing both kinds of masses plotted with the a $h^{-2}$ dependence \citep[e.g.][]{Kitzbichler2007}, or both with a $h^{-1}$ dependence \citep[e.g.][]{Bower2012}. We can only presume that they've multiplied (divided) their model (observed) masses by an additional power of $h$ to make them consistent. This is an unusual thing to do, however, as little $h$ is a measurement dependent uncertainty, and theoretical vs. observed masses are obtained through very different methods.

$\bullet$ On the topic of observed stellar masses, it's not uncommon to see an x-axis labeled with either $M_{\mathrm{stars}}$ $h^{-2}\,M_\odot$ or $M_{\mathrm{stars}}$ $/\ h^{-2}\,M_\odot$ (from Case~4). In both presentations the same meaning is usually implied: that the units of mass are $M_\odot$ and that mass has a $h^{-2}$ dependence. However it's easy to see that they are actually mathematically different; one multiplies the numerical part of the result by $h^{-2}$, while the other divides. Taken literally, you will get different results when a particular little $h$ cosmology is applied.

$\bullet$ Another such example is the presentation of galaxy magnitude. The little $h$ dependence for magnitude is commonly written (taking the $K$-band as an example) ``$M_K - 5\log(h) = \mathrm{number}$''. However this is inconsistent with the common labeling of mass discussed above (and almost all other properties): with magnitude the $h$ dependence is placed on the side with the property, whereas with mass it's placed with the numerical value of the property. Thus, to convert between $h$ cosmologies (see Section~\ref{sec:convert} below) one needs to know to treat magnitudes differently to mass.

$\bullet$ To give a final example for this section, even the author of the current paper somehow managed publish a quasar luminosity--halo mass relation where the x- and y-axes had assumed \emph{different} little $h$ cosmologies! See Figure~1 of \citealt{Croton2009} for a smile.

{\renewcommand{\arraystretch}{1.5}
\begin{table*}
\caption{A simple chart for quickly converting a property with numerical value N from a $h=1$ (or $h$'less) cosmology to $h=0.7$, close to the currently favoured value and that used in Figure~\ref{fig:fracdiff}. Seven common little $h$ scalings are shown. To go in the other direction, simply divide instead of multiply, or add instead of subtract.}
\begin{small}
\begin{center}
% \begin{tabular}{@{}l|c|c|c|c|c|c|c@{}}
\begin{tabular}{@{}l|ccccccc@{}}
\hline
  & N $h$ [units] & N $h^2$ [units] & N $h^3$ [units] & N $h^{-1}$ [units] & N $h^{-2}$ [units] & N $h^{-3}$ [units] & N$+5\log$($h$) \\
  % & N $h$ & N $h^2$ & N $h^3$ & N $h^{-1}$ & N $h^{-2}$ & N $h^{-3}$ & N $+5\log$($h$) \\
\hline 
 % & & & & & & & \\
1.0 $\Rightarrow$ 0.7 & $\times$0.700 & $\times$0.490 & $\times$0.343 & $\times$1.429 & $\times$2.041 & $\times$2.915 & $-$0.775 \\
 % & & & & & & & \\
 \hline
\end{tabular}
\end{center}
\end{small}
\label{tab:convert}
\end{table*}
{\renewcommand{\arraystretch}{1.0}

\section{Converting Between Different Hubble Parameter Values}
\label{sec:convert}

Let's say you understand all of the above, and have two sets of data that you'd like to use in your paper. Perhaps one you've collected yourself and the other has been taken from the literature. Blindly comparing the numerical values in each dataset will be lead to problems if your data has assumed one of the little $h$ cases listed in Section~\ref{sec:cases}, and the literature another. 

For example, if you're comparing the K-band magnitudes of galaxies, you may have output your data assuming Case~4, i.e. with the $h$ dependencies factored out. For the sake of argument, let us assume that the literature results you're comparing with have instead assumed $h=0.7$, as per Case~1-3. So although you may be able to find two galaxies (one from each dataset) that appear to be numerically equivalent, they of course aren't. Since little $h$ typically manifests in galaxy absolute magnitudes as $-5\log_{10}(h)$, any two such galaxies actually have a magnitude difference of $5\log_{10}(0.7)=-0.77$.

So how would you go about renormalising your galaxy magnitudes to the published $h$ values? It's simple: since little $h$ is just a number that has been factored out, replace all little $h$'s with the $h$ value you'd like to assume and evaluate. This goes for any property where little $h$ plays a part. Once the evaluation has been done, that property is then numerically correct for a universe where the Hubble constant equals that value. 

Of course the same basic rules of mathematics apply when converting between different little $h$ cosmologies, from $h=0.70$ to $h=0.73$ say. As long as you know how little $h$ presents for each property of interest you can easily and systematically reverse then reapply any $h$ value. This is one key reason why keeping the little $h$'s visible can be particularly valuable. As a reference, in Table~\ref{tab:convert} we show how property values will change when moving from $h=1$ (or $h$'less) to the commonly used value of $h=0.7$.

Figure~\ref{fig:fracdiff} shows the fractional change in little $h$ away from a value of $0.7$, when $h$ is rescaled to anywhere between $h=0.60$ and $h=0.90$, a range that brackets the currently favoured estimates (as indicated by the shaded regions). The three lines mark this difference for $h$, $h^2$ and $h^3$ dependencies. With volume measurements ($h^3$) for example, there will be a shift of over $30\%$ in a derived number between using the 2013 Planck $H_0$ value (centered on $h=0.67$) and that measured from SZ clusters (centered on $h=0.77$). When properties are expressed with the $h$'s factored out (Case~4 above), which is numerically equivalent to assuming $h=1.0$, the discrepancy can be as large as $70\%$ from Planck.

\section{Comparing Observations With Theory} 
\label{sec:theory}

If you're not using (or interested in understanding) theory-related data then you can probably skip this section. For those who do (and are), the whole little $h$ ambiguity rises to another level when you begin comparing models with observations (which is one of the primary uses of theory data, right?). 

To start with, lets assume you've obtained a popular galaxy formation model so you can over-plot its predictions against some of your own results. When using someone else's data, the first step is to understand how it was generated and the units. In particular, where appropriate one must determine which of the above four cases (or other) apply to the assumed little $h$ so that the correct conversions can be made to enable an apples-to-apples (a.k.a. fair theory-to-observation) comparison.

On the other hand, you may instead want to build your own model. Such models often must be calibrated, and this is where a reference set of observations are employed. Let's do this as an exercise, taking the standard practice of using the observed stellar mass function to calibrate the efficiency of the various model parameters. However beware! As mentioned in Section~\ref{sec:examples}, mass in numerical data typically carries a $h^{-1}$ dependence, in contrast to the $h^{-2}$ dependence often found in observations (i.e. Section~\ref{sec:measure}). The ``one less power of $h$'' comes from the way mass arises in dark matter simulations.

\begin{figure*}
\begin{center}
%gs -q -dNOCACHE -dNOPAUSE -dBATCH -dSAFER -sDEVICE=epswrite -sOutputFile=output.eps input.pdf
\includegraphics[scale=0.8, bb = 30 96 535 310, clip]{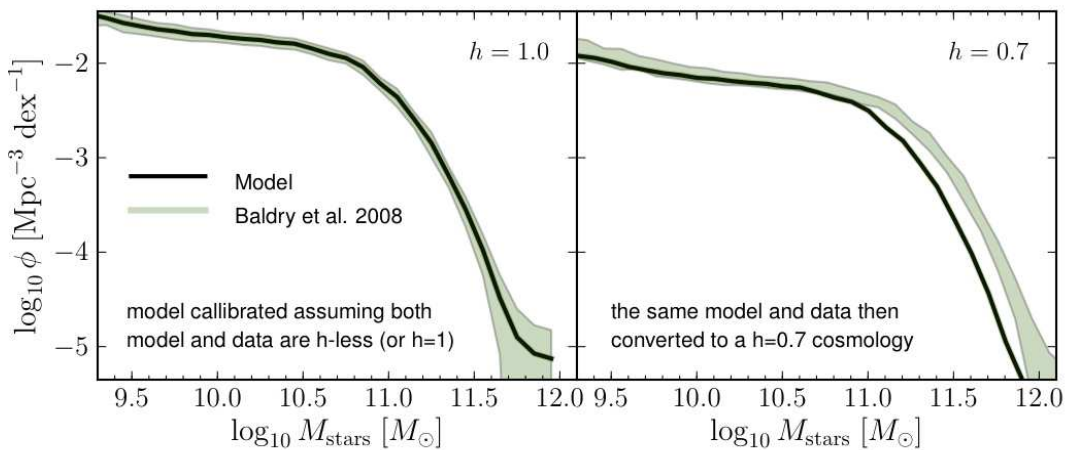}
\caption{We highlight one of the potential pitfalls when calibrating models against observations if you don't first assume a value for little $h$. In the left panel we show the stellar mass function of galaxies at $z=0$, where all $h$'s have been explicitly separated out (Case~4 of Section~\ref{sec:cases}), numerically equivalent to a universe where $h=1.0$ (and hence marked this way). Here the model (solid line) has been calibrated perfectly against the data \citep[shaded region;][]{Baldry2008}, as is commonly done. In the right panel we then update both model and data for a Universe where h=0.7 (close to the actual value). Notice that the good agreement has been lost. This is due to the different ways in which the Hubble constant manifests in these different data sets, as discussed in Section~\ref{sec:theory}.
}
\label{fig:callibration}
\end{center}
\end{figure*}

To see this, consider an expanding universe where Hubble's Law reigns. As discussed previously, distance carries a $h^{-1}$ dependence, and in fact, within the numerics of an N-body simulation all distances typically have such a scaling. Now, the masses of simulated objects are usually calculated dynamically, expressed mathematically by
\begin{equation}
% m \propto \frac{v^{3}}{H_0} ~.
\frac{G\,m}{r}\approx \sigma^{2} ~,
\end{equation}
where $G$ is the gravitational constant, and $m$ is the mass contained within a radius $r$ and supported against gravity by a velocity dispersion $\sigma$. Since $r$ is simply a distance, when masses are determined -- i.e. $m \propto \sigma^2 \ r$ -- they pick up an inverse $h$ dependence by construction. 

The differing powers of $h$ between simulation and observation must be accounted for before proceeding, and for those unclear about what little $h$ actually represents this can be a dangerous trap. For example, which set of properties should be converted, the observations or the model, and how exactly? One may think that the model properties must all be made to have exactly the same factors of $h$ as the observations, e.g. by multiplying all model masses by an additional $h$ (see one of the real-world examples in Section~\ref{sec:examples}). But this would be a grave mistake of course. On the other hand, a tempting compromise is often to explicitly factor out the $h$'s (i.e. Case~4 above) delaying a determination of the full numerical values of each property, perhaps until one can say with greater precision what the Hubble constant actually is. 

Let's do exactly this second suggestion and calibrate our model galaxy stellar mass function against its observational equivalent, in both cases keeping the (different) little $h$'s factored out. We plot the result of such an exercise with the solid line in the left panel of Figure~\ref{fig:callibration} and compare it to observational data marked by the shaded region \citep{Baldry2008}. Note the good match of the model is due to a precise calibration against the data. Note as well that this is typically how models have been historically calibrated and presented (see \citealt{Croton2006} and \citealt{Bower2006} for two popular and well cited examples).

Now let's assume that a new set of cosmological measurements are published locking the value of little $h$ down to good enough precision to be applied to our results. Remember that, as emphasised earlier, little $h$ is not a unit; it is part of the numerical value of a measurement, just an uncertain part that we were able to factor out. Once we know what it is we probably should use it.

So what will taking this new little $h$ value do to our well-tuned result? We can expect there to be a renormalisation along the y-axis, as volume has a $h^{-3}$ dependence for both model and data. Along the x-axis, however, the model will be shifted by one power of h, while the data shifted by two. The right panel of Figure~\ref{fig:callibration} shows this result for $h=0.7$. Note the good calibration is now gone. So, for this particular galaxy formation model, was the original calibration good or bad? 

The lesson is to know what little $h$ is and how it manifests in both the observations and theory, before the data is used together. Our general advice for modellers and simulators is to always work with a $H_0$ value as close to the best measured value at the time, while clearly stating the scalings for each property. In other words, do not factor out little $h$ under the guise of convenience; this is not accurate and can lead to problems later on.

Finally, when it comes to the Hubble constant and simulations, it is important to remember that N-body and pure adiabatic hydrodynamic simulations are (generally) completely scale-free, in that little $h$ can be factored out of all properties and the simulation scaled to any $h$ value in post-processing. This of course also holds true for the semi-analytic and halo occupation distribution models that are constructed on top of N-body simulation merger trees. However, for more sophisticated hydrodynamic simulations, where absolute time-scales, distances, temperatures, etc need to be established to model processes like cooling, star formation, and supernova feedback, such little $h$ scaling breaks and one can only work with the Hubble constant assumed when the simulation was originally run.

\section{Summary: Recommendations For Using And Expressing Little h}
\label{sec:summary}

On the surface, the use of little $h$ when quoting the values of observed or simulated galaxy properties appears simple. However, in practice it can get confusing due to the different ways the Hubble constant can manifest in data, and the different ways that authors present their results. 

In this paper we provide an introduction to the origin of the Hubble constant and the definition of little $h$, describe how little $h$ arises in the measurement of galaxy properties (notably for survey data), and highlight four general ways in which little $h$ is commonly expressed in the literature. We then walk through the method to convert between galaxy properties that have been expressed using different values of the Hubble constant. This is notably tricky when comparing observed and simulated results, where the same galaxy property can have a different little $h$ dependence.

Our take-home message is this: First, the clearest way to express your results is to state the $h$ scaling of each property at the beginning of your paper, then evaluate all properties assuming your best guess for the actual $h$ cosmology of the Universe, e.g. $h=0.7$. Once that's done for all results presented (measured in as many different ways as you like), each numerical value will be the actual value (assuming that cosmology) and can be compared with any other having the same units. 

Second, the most sensible, less error prone approach to little $h$ is to treat it like any other uncertainty in the data. The Hubble constant is a relic of bygone times. It is one of the least uncertain parameters in astrophysics, and the case for continuing to single it out as an independent parameter is weak. 

Third, we emphasise that little $h$ is not a unit. Units are physical quantities, like $M_\odot$, Mpc, and km/s. It is best to be explicit about this. If you feel compelled to display little $h$ when presenting each result, then separate out the units of the property from the combination of the numerical value and little $h$. For example, masses should be written $M =  10^{12} h^{-1} M_\odot$, not $M = 10^{12} M_\odot\, h^{-1}$ or $10^{12} M_\odot/h$. The latter two imply the $h$ is coupled to $M_\odot$, which is not true.

To conclude, in Table~\ref{tab:scalings} we provide a list of common galaxy properties, their little $h$ dependencies, and popular units to aid the uninitiated data user. Our key points are summarised below in a ``cheat sheet'' providing nine rules for dealing with little $h$. Follow these and we will all live happier lives as a result.

\begin{itemize}

\item[1] Little $h$ is NOT a unit. $h$ expresses an unknown part of the numerical value of a property. Units are $L_\odot$, Mpc, etc. $h$ is a number.

\item[2] Little $h$ manifests in a galaxy property due to the method of measurement. For example, to measure the luminosity of a galaxy its luminosity distance must be known, and cosmological distances carry a $h$ dependence. 

\item[3] In general, when $h$ is presented in a property you know that the $h$ dependence has been factored out, with the $h$ scaling explicitly shown.

\item[4] To put the property into a particular $h$ cosmology, replace the $h$ in the property with the desired $h$ value and evaluate.

\item[5] To change from an already assumed $h$ to a new value, reverse the above by doing the opposite with the assumed $h$. You can then substitute the new $h$ value in.

\item[6] The terminology ``$h=1.0$'' and ``$h$'s factored out'' are often used interchangeably in the literature. Mathematically, the numerical value of a property will be identical. But in reality they represent different things. The first assumes a particular Hubble constant. The second assumes none and lets you decide at a later date.

\item[7] Sometimes the same property can be measured in different ways that have different dependencies on $h$, e.g. mass estimated from luminosity ($h^{-2} M_\odot$), from dynamics ($h^{-1} M_\odot$), or even in non-cosmological ways (e.g. reverberation mapping) that have no $h$ dependence ($M_\odot$). In such cases, as you change little $h$ the different determinations of the property will scale differently. 

\item[8] When you want to compare the same property that was obtained two different ways having different dependencies on $h$ (e.g. theoretical and observed masses), DO NOT transform one to match the $h$ scaling of the other. Instead, choose a $h$ value and convert both independently to that cosmology using point 4 above. Only then can they be meaningfully compared.

\item[9] Always clearly state what you have assumed with regards to little $h$ at the start of your paper. For the love of God!

\end{itemize}

% \section*{Appendix}

{\renewcommand{\arraystretch}{1.2}
\begin{table*}
\caption{An (incomplete) list of common galaxy properties, their little $h$ scalings, and units.}
\begin{small}
\begin{center}
\begin{tabular}{@{}lcll@{}}
\hline
 Property & Common $h$ Scaling & Common Units & Notes \\
\hline 
\\
 Time, Age 										& $h^{-1}$ 				& yr, Myr, Gyr 						& Can also have no $h$ scaling \\
 & & & e.g. SN1A decay rates \\
\\
 Distance 										& $h^{-1}$ 				& kpc, Mpc 								& \\
 Area 												& $h^{-2}$ 				& kpc$^{2}$, Mpc$^{2}$ 		& \\
 Volume 											& $h^{-3}$ 				& kpc$^{3}$, Mpc$^{3}$ 		& \\
\\
 Mass (from luminosity) 			& $h^{-2}$ 				& M$_\odot$ 							& Common in observations \\
 Mass (from dynamics) 				& $h^{-1}$ 				& M$_\odot$ 							& Common in simulations \\
 Mass (direct measure) 				& none						& M$_\odot$ 							& \\
\\
 Luminosity 									& $h^{-2}$ 				& L$_\odot$ 							& \\ 
 Absolute magnitude 					& $+5\log$($h$) 	&  												& Note: mag$-5\log$($h$)=N \\
 & & & $\ \Rightarrow$ mag=N$+5\log$($h$) \\
 Apparent magnitude 					& none 						& 												& \\
 Surface brightness 					& none 						& mag arcsec$^{-2}$ 			& Physical units may \\
 & & & collect a $h$ scaling \\
\\
 Velocity (models/simulations) & none						& km s$^{-1}$ 						& $h$ scalings cancel (see above) \\
 Velocity (line-widths)				& none						& km s$^{-1}$ 						& \\
\\
 SFR (models/simulations)			& none 						& M$_\odot$ yr$^{-1}$ 		& $h$ scalings cancel (see above) \\
 SFR (from luminosity) 				& $h^{-2}$ 				& M$_\odot$ yr$^{-1}$ 		& \\
\\
 Temperature 									& 		 						& K 											&	The $h$ scalings here can \\
 Power 												& 				 				& erg s$^{-1}$ 						& be highly dependent on \\
 Pressure 										& 	 							& K cm$^{-3}$ 						& the measurement method \\
\hline
\end{tabular}
\end{center}
\end{small}
\label{tab:scalings}
\end{table*}
{\renewcommand{\arraystretch}{1.0}

\section*{Acknowledgments}
Many thanks to Alan Duffy, Thibault Garel, Karl Glazebrook, David Hogg, Alexander Knebe, John Peacock, Greg Poole, Jeremy Mould, Simon Mutch, Genevieve Shattow, and Adam Stevens for their valuable comments at various stages of this work. Also to the anonymous referee for his/her constructive report. Special thanks in particular to Alan Duffy and Thibault Garel for the numerous discussions which helped highlight many of the issues presented here and their resolution. The author wishes to acknowledge receipt of a QEII Fellowship by the Australian Research Council.

%\end{multicols}

\end{document}